\documentstyle[amssymb,prl,aps,epsf]{revtex}

\begin{document}
\draft
\title{Observation of Coherent Charge-State Mixing in Asymmetric Bloch Transistors }
\author{Daniel J. Flees and J.E. Lukens}
\address{Department of Physics and Astronomy, University at Stony Brook, Stony\\
Brook, NY 11794-3800}
\author{Siyuan Han}
\address{Department of Physics and Astronomy, University of Kansas, \\
Lawrence, KS 66045}
\maketitle

\begin{abstract}
The reduced switching current of an excited energy-band in a Bloch
transistor can be used to detect microwave-induced interband transitions.
Spectroscopic measurements of the band-gap are performed by mapping the
frequency dependence of the excitation threshold where the microwave photon
energy and band-gap are equal. This excitation threshold also provides a
probe of charge-state mixing on the island of the transistor. Any asymmetry
in the junctions of the transistor makes the coherent mixing of
charge-states appear as a finite gap-splitting at the electrostatic
degeneracy point between states differing by one excess Cooper-pair on the
island. The measured gap-splitting in an asymmetric transistor clearly
demonstrates the effect.
\end{abstract}

\pacs{}

The development of nanometer scale tunnel junctions has allowed the control
and study of single-electron tunneling in mesoscopic structures\cite{SCT-2} 
\cite{SCT-1}\cite{FRENCH-1}. Much of this work has been done using Al/AlO$_x$%
/Al tunnel junctions with a magnetic field employed to suppress
superconductivity. Investigation of such systems in the superconducting
state permits elimination of the dissipation associated with electrons
tunneling across a finite voltage drop. These experiments depend heavily
upon adequate microwave shielding \cite{SMALLJJ-1}\cite{FILTER-2}\cite
{FILTER-3}, since the coherent effects of interest are highly noise
sensitive. This added difficulty has plagued experimental efforts until
recently. However, with proper noise isolation coherent charge-tunneling
effects can be observed\cite{NAKAMURA}\cite{COMMENT-0}. One of the most
commonly studied systems is the superconducting version of the
single-electron transistor, or Bloch transistor \cite{SCT-2}\cite{JQP-1}\cite
{geerligs}\cite{EOP-8}\cite{SSET-2}\cite{EOP-4}\cite{DJFPRL-1}. It consists
of a pair of small Josephson junctions coupled by a superconducting island (%
{\rm Fig. 1a}). An electrostatically coupled gate electrode controls the
Coulomb blockade against charge tunneling onto the island. The Hamiltonian
describing this system (neglecting interaction with the external
environment) can be written,

\begin{equation}  \label{1}
H_0=E_c(n-Q)^2-\sum\limits_{i=1,2}E_{Ji}\text{cos}(\varphi _i),
\end{equation}
where E$_c$ = e$^2$/2C$_\Sigma $ is the charging energy of the island and E$%
_{J1}$, E$_{J2}$, $\varphi _1$,$\varphi _2$ are the Josephson coupling
energies and phases of the two junctions. The polarization or quasicharge Q
on the island is given by Q = C$_g$V$_g$/e + Q$^{^{\prime }}$, where C$_g$,V$%
_g$ are the gate capacitance and voltage and Q$^{^{\prime }}$ is
polarization induced by background charge. If the transistor is symmetric, E$%
_{J1}$ = E$_{J2}$ $\equiv $ E$_J$, the Hamiltonian (Eq. 1) can be simplified
in a more natural set of variables, $\varphi \equiv $ $\varphi _1$ + $%
\varphi _2$ and $\theta \equiv $ ($\varphi _1$ - $\varphi _2$)/2, giving,

\begin{equation}  \label{2}
H_0=E_c(n-Q)^2-2E_Jcos(\varphi /2)cos(\theta ).
\end{equation}
This Hamiltonian consists of the charging energy associated with n excess
electrons on the island and an effective coherent coupling, 2E$_J$cos($%
\varphi $/2), between charge states separated by a single Cooper-pair
tunneling. It contains the basic prescription for studying the competition
between charging which results in discrete energy states and coherent
charge-state mixing which modifies the energy levels. This competition is
characterized by the ratio $\alpha \equiv $ E$_J$/E$_c$. The Hamiltonian
(Eq. 2) can be diagonalized in a subspace of charge states 
\mbox{$\vert$}%
n%
\mbox{$>$}%
where n = 0,$\pm $2,..., or n = $\pm $1,$\pm $3,..., depending on the parity
of the ground state. At a fixed value of the quasicharge Q, the energy
eigenvalues form a unique set of energy bands E$_m$($\varphi $)%
\mbox{$\vert$}%
$_Q$, each of which is analogous to a Josephson washboard potential.
Consequently, each band can support a maximum supercurrent given by,

\begin{equation}  \label{3}
I_{c,m}(Q)=(2e/\hbar )max\{\partial E_m/\partial \varphi |_Q\}.
\end{equation}
Sweeping the voltage applied to the gate of the transistor moves the system
along the quasicharge axis, producing a characteristic 2e-periodic
modulation in the critical currents of the bands and the energy-gap between
them. Since the minimum energy separation between the ground and first
excited band (at fixed Q) occurs at $\varphi $ = $\pi $, it is particularly
instructive to consider the energy states E$_m$(Q)%
\mbox{$\vert$}%
$_{\varphi =\pi }$ ({\rm Fig. 1b}). The coherent coupling term is absent in
Eq. 2 at $\varphi $ = $\pi $ and the energy states correspond to those of a
superconducting single-electron box (S-SEB) \cite{BOX-1} in the limit E$%
_J\rightarrow 0$ \cite{EOP-2}. In the reduced zone picture (-1 $\leq $ Q $%
\leq $ 1), as the gate voltage sweeps the quasicharge, the system follows
the lowest energy charging parabola, E = E$_c$(n-Q)$^2$, with fixed excess
charge n on the island. At the zone edge (Q = 1), there is a net charge
tunneling, incrementing the average charge on the island and taking the
system back to the opposite zone edge (Q = -1). By tracing the lowest energy
charging parabolas within a band, the Q dependence of the average charge on
the island at $\varphi $ = $\pi $ is obtained ({\rm Figs. 1c -} {\rm 1d}).
Clearly, when the coherent coupling is not present, the energy-gap in the
reduced zone is just the difference between the charging energies of states
differing by one Cooper-pair on the island, E$_G=$ 4E$_c$(1-%
\mbox{$\vert$}%
Q%
\mbox{$\vert$}%
). In particular, the two lowest bands are degenerate at the zone edge.

The situation becomes more interesting when $\varphi $ $\neq $ $\pi $ and
the coherent coupling term is present. In this case, coherent mixing of
charge states occurs and the degeneracy at the band edges (Q = $\pm $1) is
lifted (dashed lines in {\rm Fig. 1b}). The effect of the mixing of charge
states is also clearly seen in the average charge on the island, where
deviations from the S-SEB (E$_J\rightarrow 0$ limit) occur when $\varphi $ $%
\neq $ $\pi $ ({\rm Figs. 1d - 1c}). Unfortunately, $\varphi $ is a
dynamical variable which would seem to prevent a systematic study of this
coherent effect \cite{COMMENT-0}. However, if the transistor has asymmetric
junctions, the degeneracy at Q = $\pm $1 is again lifted because in this
case the coherent coupling terms do not cancel at $\varphi $ = $\pi $. The
general expression for the energy gap in an asymmetrical transistor is

\begin{equation}  \label{4}
E_G=\sqrt{(E_{J1}-E_{J2})^2+(4E_c(1-|Q|))^2}.
\end{equation}

Measurements of the band-gap between the lowest bands can be made using the
fact that the excited band has a lower critical current than the ground
band. Microwave induced excitations resulting in a reduction of the
switching current will occur when E$_G$ $\leq $ $\hbar \omega $. According
to Eq. 4, for each frequency $\omega $, there is a threshold charge $Q_{th},$
set by the condition E$_G$($Q_{th}$) = $\hbar \omega $, beyond which no
transitions can be excited because the band-gap is larger than a single
photon's energy\cite{COMMENT-1}. Measurement of the frequency dependence of $%
Q_{th}$ constitutes a spectroscopic evaluation of the energy band-gap.
Further, at $Q_{th}$ the photon energy is matched to the minimum energy
separation which occurs at $\varphi $ = $\pi $, so the system can only be
making transitions from this phase region. Thus measurements of the
threshold charge probe the energy states of the transistor where it is
equivalent to the S-SEB ({\rm Fig 1b}). While a symmetric transistor is
equivalent to the limit E$_J\rightarrow 0$, the asymmetric transistor is
analogous to a S-SEB with a coherent coupling given by E$_J=|E_{J1}-E_{J2}|$.

Previous measurements of the band-gap in Bloch transistors were performed on
nominally symmetric devices \cite{DJFPRL-1}. Figure 2 shows the effects of
microwaves on the modulation of the supercurrent for such a device. The
critical currents of the bands are calculated using Eq. 3 once the
Hamiltonian (Eq. 2) has been diagonalized and the energy bands E$_m$(Q,$%
\varphi $) extracted. The fit to the data is then achieved using an
empirically measured relation between the critical currents (I$_c$) of
Josephson junctions \cite{COMMENT-2} and the measured switching currents (I$%
_s$). In the low damping limit, there is a theoretical prediction that I$_s$ 
$\propto $ I$_c^{3/2}$ \cite{EOP-4}. Empirically we find a nearly
temperature independent exponent, in the range 1.62 to 1.64 between 30 and
100 mK. The vertical bars in Fig. 2 show the expected position of $Q_{th}$
at 28 GHz for the band-gap between the lowest energy bands. The charging
energy derived from the band-gap measurement was 0.52 K and the Josephson
coupling inferred from the normal state resistance (27.5 k$\Omega $) and the
superconducting gap ($\Delta $ $\sim $ 200 $\mu $V) was 0.54 K, giving the
estimate $\alpha =1$. The best fit using the relation I$_s$ $\propto $ I$%
_c^{1.63}$ gives $\alpha $ $\sim $ 0.9, in reasonable agreement with the
estimated value. The microwave power level for the data presented in {\rm %
Figure 2} is rather high, causing a small switching current suppression even
above $Q_{th}$. The dip at Q = 0 may be attributed to multiphoton absorption
and is suggestive of the presence of the second excited band which is
degenerate with the first excited band at this point.

The limited resolution of the measurements near the degeneracy points (Q = $%
\pm $ 1) makes it impossible to detect any gap-splitting caused by coherent
charge-state mixing in nominally symmetric transistors where only a small
asymmetry is expected due to variance in the fabricated junction sizes. In
order to confirm the band-gap (Eq. 4), and observe the gap-splitting
produced by coherent charge-state mixing, the asymmetry in the junctions can
be made large enough so that ($E_{J1}-E_{J2}$)/$E_c$ $\sim $ 1. Then the
effect of the gap-splitting should be apparent even far from the degeneracy
points at Q = $\pm $1, making it easily measureable. {\rm Figure 3} shows
microwave band-gap measurements at two different frequencies for a
transistor with a large junction asymmetry by design. The total normal state
resistance of the device is R$_n=$ 18.3 k$\Omega $. The junction asymmetry
can be determined from the asymmetry in the slopes of the voltage modulation
produced by the Q dependence of the Coulomb blockade in the normal state\cite
{COMMENT-3}. The measured asymmetry for this transistor is 1.9 $\pm $ 0.2
although the designed junction sizes were 0.01 $\mu $m$^2$ and 0.025 $\mu $m$%
^2$. Thus the estimated Josephson coupling energies are E$_{J1}=$ 1.19 K and
E$_{J2}=$ 0.62 K. The charging energy is not easily measurable for
transistors with R$_n\leq $ R$_k=h/e^2=25.8$ k$\Omega $, so it is treated as
a fitting parameter to be compared with estimates based on the designed
junction sizes and a previously measured specific capacitance C$_{\Box }=$
43fF/$\mu $m$^2$ \cite{DJFPRL-1}. {\rm Figure 4} shows the results of the
band-gap measurement. The error bar at Q = 1 is not determined by a direct
measurement of $Q_{th}$, but rather by decreasing the microwave frequency
until no Q dependent switching current suppression is seen at any input
power level up to that which causes Q independent suppression. The
gap-splitting at Q = 1 due to coherent charge-state mixing is evident and
the fit to Eq. 4 is reasonably good using the inferred Josephson coupling
energies and a charging energy of 0.6 K. The fit charging energy compares
well with the estimate based on the designed junction sizes (0.62 K) even
though the junction asymmetry indicates that the ratio of junction areas
differs from the designed value by $\sim $ 25 \%. This probably indicates
that most of the size deviation has occurred in the smaller junction.

In summary, microwave induced interband transitions provide a spectroscopic
probe of the band-gap in Bloch transistors. Transitions which occur at the
threshold charge where the band-gap and photon energy are equal can only
take place in a limited phase region near $\varphi =\pi $, where the
behavior of the transistor is simply related to mixing of charge states on
the island. Symmetric transistors which have no coherent coupling in this
region, have a degeneracy in the energy bands. In asymmetric transistors the
coherent coupling between charge-states differing by one excess Cooper-Pair
on the island, lifts the degeneracy. The data presented clearly show this
coherence induced gap-splitting and are in quantitative agreement with the
expected band-gap over the measured range of quasicharge Q.

The authors wish to acknowledge the contributions of K. K. Likharev and D.
V. Averin to our understanding of the Bloch transistor. This work has been
supported in part by the AFOSR.

\bibliographystyle{c:/sciword/bibtex/prsty}
\bibliography{c:/sciword/bibtex/set}

\begin{thebibliography}{10}
\bibitem{COMMENT-1}
  Multiple photon absorption can cause transistions above Q$_{th}$. For example,
  a threshold charge corresponding to a band-gap of 40 GHz sometimes appears
  when applying high power levels of 20 GHz radiation, even though poor
  coupling and attenuation prevent excitation of transitions when directly
  applying 40 GHz radiation.
\end{thebibliography}

\newpage 

\begin{figure}[tbp]
\caption{a) Schematic of a Bloch transistor which is characterized by the
Josephson coupling energies of the junctions (E$_{J1}$, E$_{J2}$) and the
charging energy of the island capacitance (C$_\Sigma $ = C$_1$ + C$_2$ + C$%
_g $). b) Energy states for a symmetic transistor at $\protect\varphi = 
\protect\pi$ (solid lines) where there is no coherent coupling and $\protect%
\varphi \neq \protect\pi$ (dashed lines) where coherent coupling lifts the
degeneracy at the reduced zone edges. c) and d) Calculated average charge on
the island of a transistor with $\protect\alpha$ = 1.5 in c) the ground band
(m=0) and d) the first excited band (m=2). }
\label{fig1}
\end{figure}

\begin{figure}[tbp]
\caption{ The effect of 28 GHz irradiation on the switching current
modulation of a nominally symmetric transistor. The raw switching data at 30
mK, with and without microwaves, are shown overlaid with the theoretical
prediction. An empirically measured relation has been used to convert the
calculated critical currents to switching currents - see text for details. }
\label{fig2}
\end{figure}

\begin{figure}[tbp]
\caption{Band-gap measurements at a) 24 GHz and b) 32 GHz. The data were
taken at 80 mK and sets with and without irradiation are shown in each
panel. The lower traces are the differences between the averaged switching
distributions of the sets (offset and scaled for clarity). The vertical bars
represent the threshold charge (Q$_{th}$) determined from several modulation
periods. }
\label{fig3}
\end{figure}

\begin{figure}[tbp]
\caption{The measured frequency dependence of the threshold charge Q$_{th}$
for an asymmetric transistor. The error bars are attributable primarily to
fluctuations in the background charge (Q$^{^{\prime }}$) and are estimated
based on measurements over several modulation periods. The solid line is a
fit to Eq. 4 using the measured junction asymmetry while the dashed line is
the expected fit for a symmetric transistor with equivalent charging energy. 
}
\label{fig4}
\end{figure}

\newpage
 
\begin{figure}[tbp]
\setlength{\unitlength}{1.0in}
\begin{center}
\begin{picture}(6.0,8.0)
\put(0.15,-0.03){\epsfxsize=4in\epsfysize=6in\epsfbox{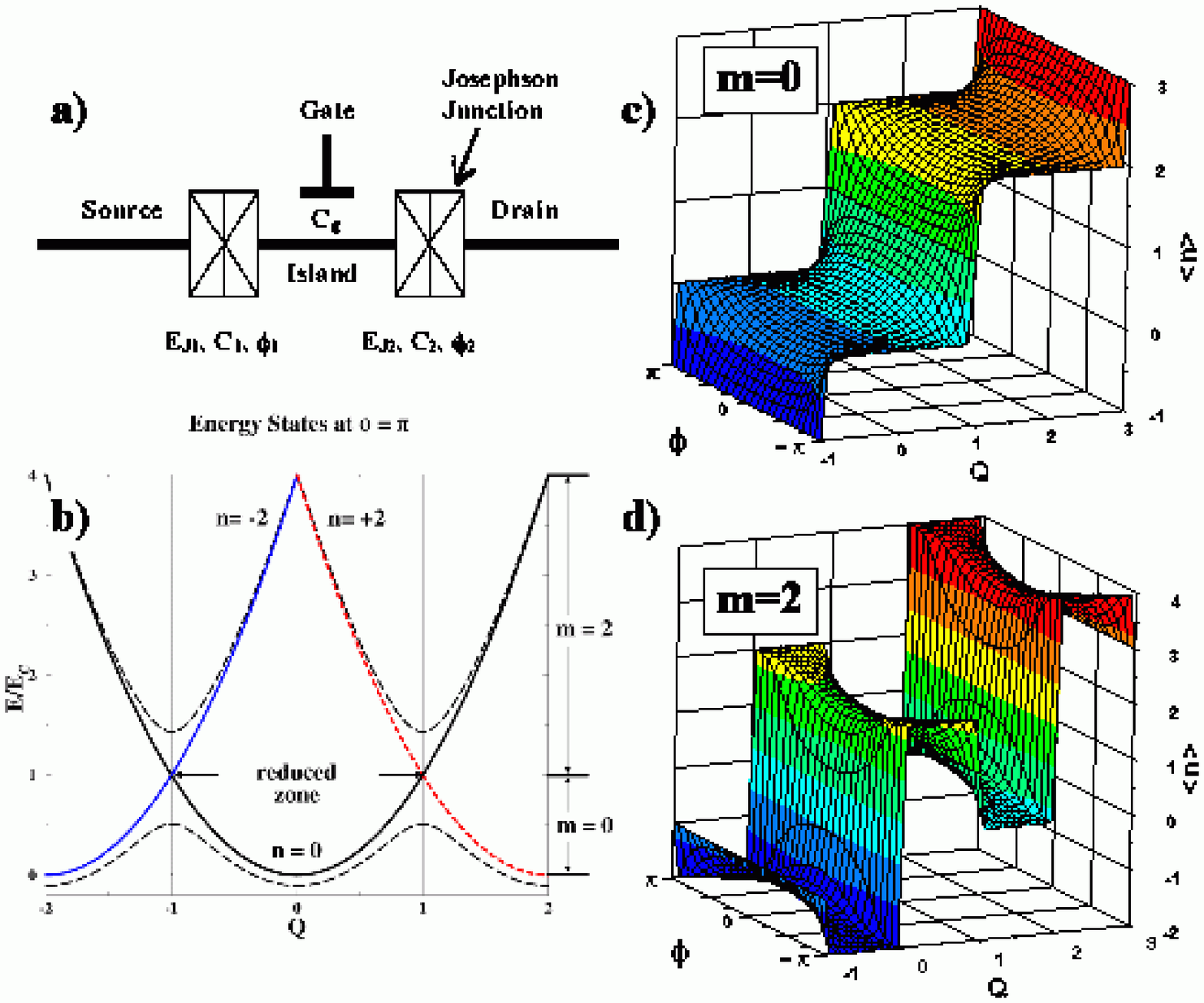}}
\end{picture}
\end{center}
\end{figure}

\begin{figure}[tbp]
\setlength{\unitlength}{1.0in}
\begin{picture}(6.0,8.0)
\put(0.15,-0.03){\epsfxsize=4in\epsfysize=6in\epsfbox{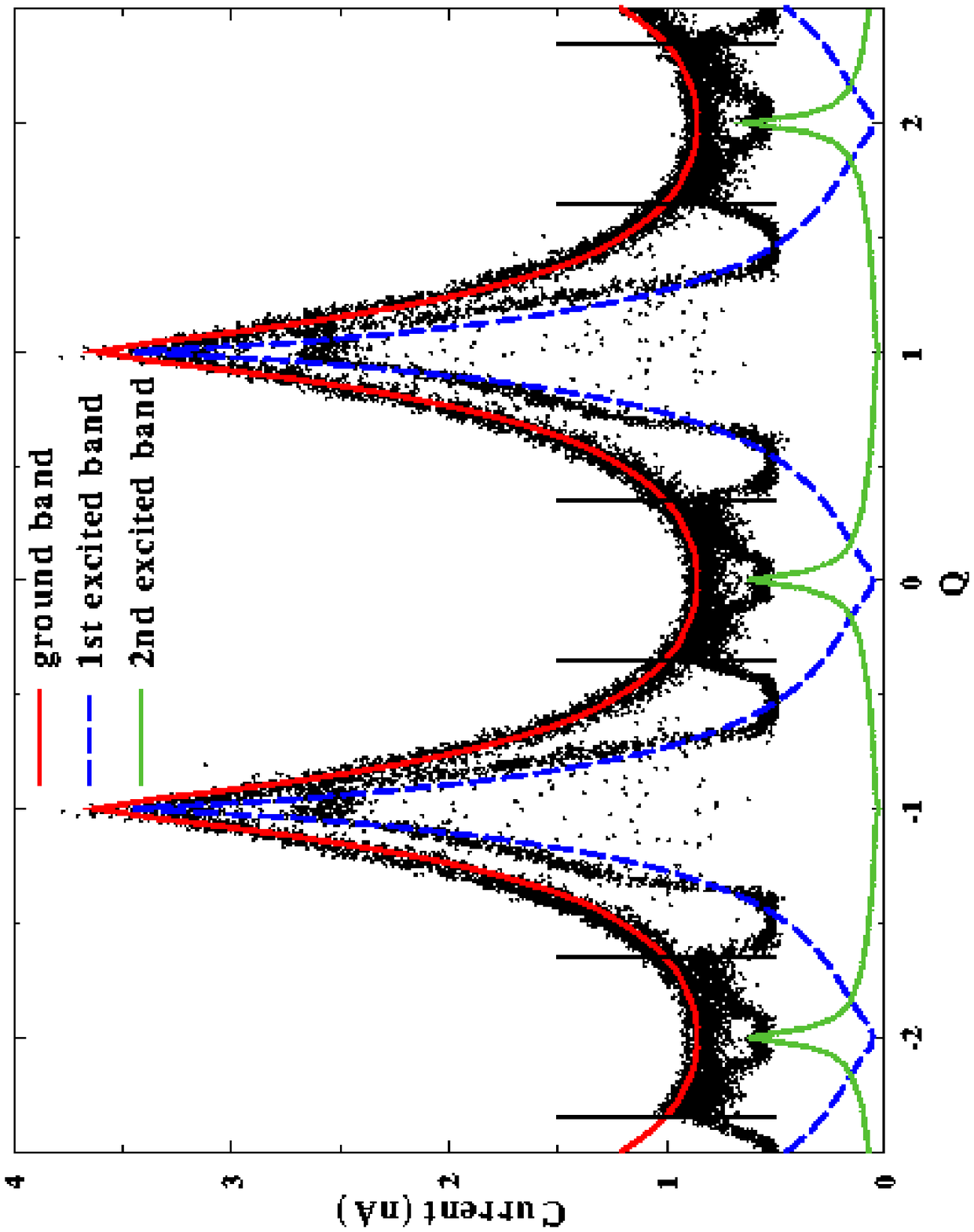}}
\end{picture}
\end{figure}

\begin{figure}[tbp]
\setlength{\unitlength}{1.0in}
\begin{picture}(1.0,8.0)
\put(0.15,-0.03){\epsfxsize=4in\epsfysize=6in\epsfbox{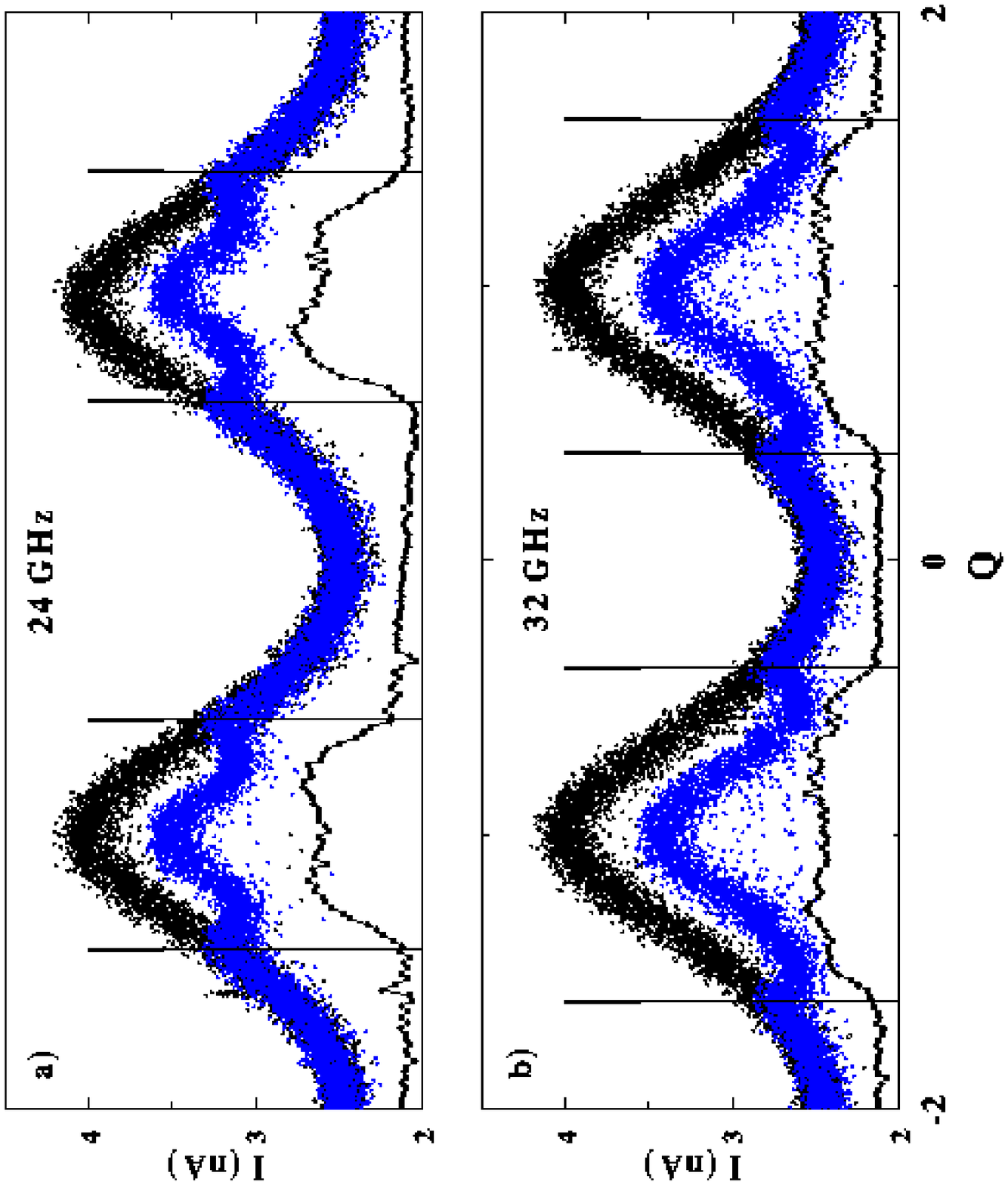}}
\end{picture}
\end{figure}

\begin{figure}[tbp]
\setlength{\unitlength}{1.0in}
\begin{picture}(1.0,8.0)
\put(0.15,-0.03){\epsfxsize=4in\epsfysize=6in\epsfbox{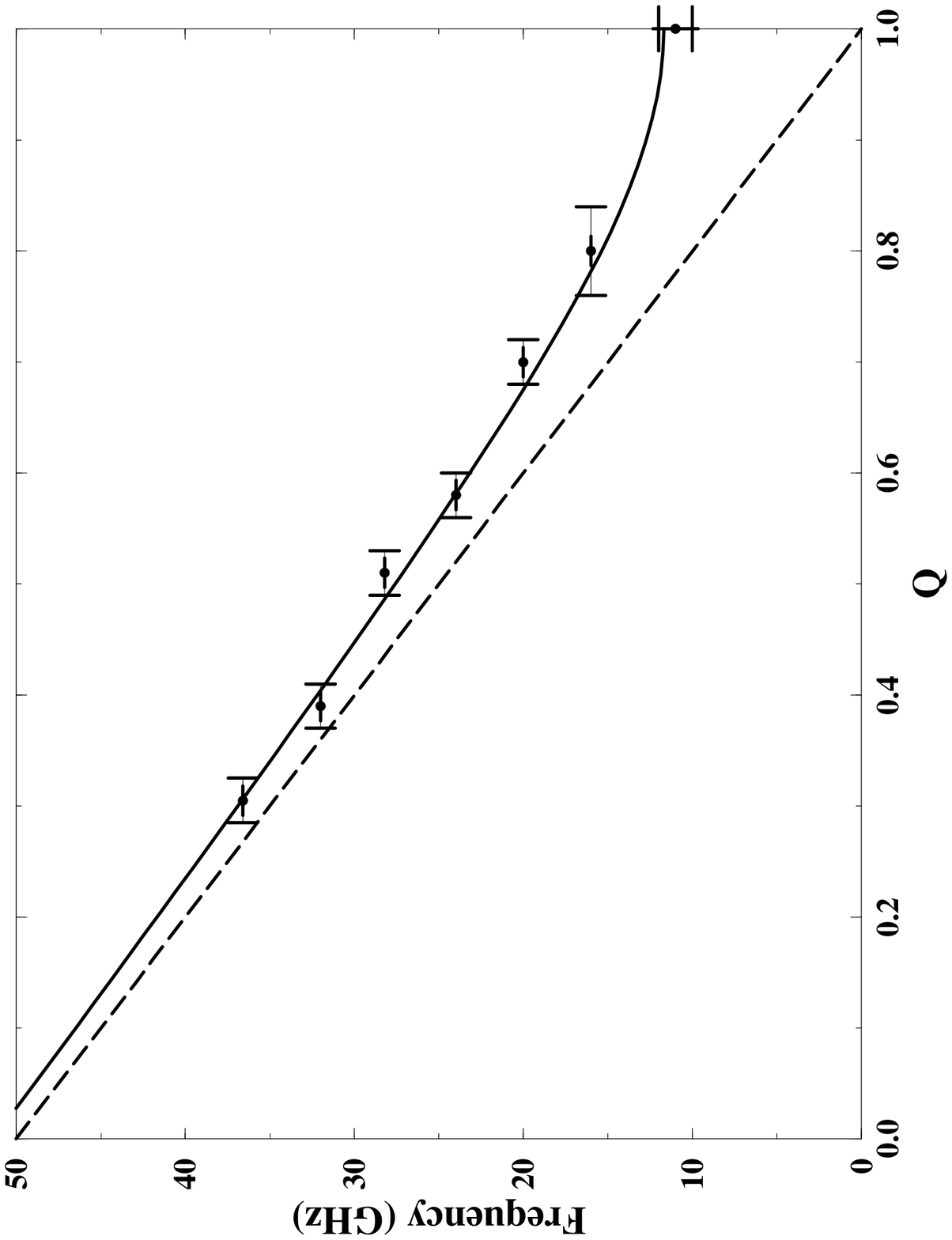}}
\end{picture}
\end{figure}

\end{document}